\documentclass[final,3p,times,twocolumn]{elsarticle}
\usepackage{epsfig}

\usepackage{amssymb}
\def\C {\chi_c}

\def\g {\gamma}
\def\J {J/\psi}
\def\Y {\Upsilon}

\def\R {{\cal R}}

\def\ktf {$k_t$-factorization }

\def\n {\nonumber}
\def\ap#1#2#3   {{\rm Ann. Phys. (NY)}       #1 (#3) #2}
\def\apj#1#2#3  {{\rm Astrophys. J.}         #1 (#3) #2}
\def\apjl#1#2#3 {{\rm Astrophys. J. Lett.}   #1 (#3) #2}
\def\app#1#2#3  {{\rm Acta. Phys. Pol.}      #1 (#3) #2}
\def\cpc#1#2#3  {{\rm Computer Phys. Comm.}  #1 (#3) #2}
\def\dum#1#2#3  {{~}                         #1 (#3) #2}
\def\epjc#1#2#3 {{\rm Eur. Phys. J. C}       #1 (#3) #2}
\def\err#1#2#3  {{\it Erratum}               #1 (#3) #2}
\def\ib#1#2#3   {{\it ibid.}                 #1 (#3) #2}
\def\jcp#1#2#3  {{\rm J. Comp. Phys.}        #1 (#3) #2}
\def\jmp#1#2#3  {{\rm J. Math. Phys.}        #1 (#3) #2}
\def\jhep#1#2#3 {{\rm JHEP}                  #1 (#3) #2}
\def\ijmp#1#2#3 {{\rm Int. J. Mod. Phys.}    #1 (#3) #2}
\def\jpg#1#2#3  {{\rm J. Phys. G.}           #1 (#3) #2}
\def\mpl#1#2#3  {{\rm Mod. Phys. Lett.}      #1 (#3) #2}
\def\nat#1#2#3  {{\rm Nature (London)}       #1 (#3) #2}
\def\ncim#1#2#3 {{\rm Nuovo Cimento}         #1 (#3) #2}
\def\nca#1#2#3  {{\rm Nuovo Cimento A}       #1 (#3) #2}
\def\ncb#1#2#3  {{\rm Nuovo Cimento B}       #1 (#3) #2}
\def\nim#1#2#3  {{\rm Nucl. Instr. Meth.}    #1 (#3) #2}
\def\np#1#2#3   {{\rm Nucl. Phys.}           #1 (#3) #2}
\def\npb#1#2#3  {{\rm Nucl. Phys. B}         #1 (#3) #2}
\def\pan#1#2#3  {{\rm Phys. At. Nuclei}      #1 (#3) #2}
\def\pl#1#2#3   {{\rm Phys. Lett.}           #1 (#3) #2}
\def\plb#1#2#3  {{\rm Phys. Lett. B}         #1 (#3) #2}
\def\prep#1#2#3 {{\rm Phys. Rep.}            #1 (#3) #2}
\def\prev#1#2#3 {{\rm Phys. Rev.}            #1 (#3) #2}
\def\prc#1#2#3  {{\rm Phys. Rev. C}          #1 (#3) #2}
\def\prd#1#2#3  {{\rm Phys. Rev. D}          #1 (#3) #2}
\def\prev#1#2#3 {{\rm Phys. Rev.}            #1 (#3) #2}
\def\prl#1#2#3  {{\rm Phys. Rev. Lett.}      #1 (#3) #2}
\def\prs#1#2#3  {{\rm Proc. Roy. Soc.}       #1 (#3) #2}
\def\ptp#1#2#3  {{\rm Prog. Theor. Phys.}    #1 (#3) #2}
\def\ps#1#2#3   {{\rm Physica Scripta}       #1 (#3) #2}
\def\rmp#1#2#3  {{\rm Rev. Mod. Phys.}       #1 (#3) #2}
\def\rpp#1#2#3  {{\rm Rep. Prog. Phys.}      #1 (#3) #2}
\def\sjnp#1#2#3 {{\rm Sov. J. Nucl. Phys.}   #1 (#3) #2}
\def\spj#1#2#3  {{\rm Sov. Phys. JETP}       #1 (#3) #2}
\def\spu#1#2#3  {{\rm Sov. Phys.-Usp.}       #1 (#3) #2}
\def\yaf#1#2#3  {{\rm Yad. Fiz.}             #1 (#3) #2}
\def\zp#1#2#3   {{\rm Zeit. Phys.}           #1 (#3) #2}
\def\zpa#1#2#3  {{\rm Zeit. Phys. A}         #1 (#3) #2}
\def\zpc#1#2#3  {{\rm Zeit. Phys. C}         #1 (#3) #2}
\def\et{{\rm et al.}}
\journal{Physics Letters B}

\begin{document}

\begin{frontmatter}

%% Title, authors and addresses

%% use the tnoteref command within \title for footnotes;
%% use the tnotetext command for theassociated footnote;
%% use the fnref command within \author or \address for footnotes;
%% use the fntext command for theassociated footnote;
%% use the corref command within \author for corresponding author footnotes;
%% use the cortext command for theassociated footnote;
%% use the ead command for the email address,
%% and the form \ead[url] for the home page:
%% \title{Title\tnoteref{label1}}
%% \tnotetext[label1]{}
%% \author{Name\corref{cor1}\fnref{label2}}
%% \ead{email address}
%% \ead[url]{home page}
%% \fntext[label2]{}
%% \cortext[cor1]{}
%% \address{Address\fnref{label3}}
%% \fntext[label3]{}

\title{Double heavy meson production through double parton scattering in hadronic
collisions}

%% use optional labels to link authors explicitly to addresses:
%% \author[label1,label2]{}
%% \address[label1]{}
%% \address[label2]{}

\author{S.P.\ Baranov,}

\address{P.N. Lebedev Institute of Physics, 
         119991 Moscow, Russia}
\author{A.M.~Snigirev, N.P. Zotov}

\address{D.V. Skobeltsyn Institute of Nuclear Physics, M.V. Lomonosov Moscow 
State University, 119991, Moscow, Russia }

\begin{abstract}
It is shown that the contribution from double parton scattering to the 
inclusive double heavy meson yield is quite comparable with the usually 
considered mechanism of their production at the LHC energy. For some 
pairs of heavy flavored quarks in the final state the double parton 
scattering will be a dominant mode of their production.
\end{abstract}

\begin{keyword}
double parton scattering \sep heavy mesons  
\PACS 12.38.-t \sep 13.85.-t \sep 13.85.Dz

%% MSC codes here, in the form: \MSC code \sep code
%% or \MSC[2008] code \sep code (2000 is the default)
\end{keyword}

\end{frontmatter}

%% \linenumbers

%% main text
%\section{Introduction}

In the last years it has become obvious that multiple parton interactions 
play an important role in hadron-hadron collisions at high energies and 
are one of the most common, yet poorly understood~\cite{workshop}, 
phenomenon at the LHC. The presence of such multiple parton interactions 
in high-energy hadronic collisions has been convincingly demonstrated by 
the AFS~\cite{AFS}, UA2~\cite{UA2}, CDF~\cite{cdf4jets,cdf}, and 
D0~\cite{D0} Collaborations, using events with the four-jets and 
$\gamma+3$-jets final states, thus providing new and complementary 
information on the proton structure. The possibility of observing two 
separate hard collisions was proposed long ago. Early theoretical 
investigations were carried out in the framework of the parton 
model~\cite{landshoff,takagi,goebel} with subsequent extension to 
perturbative QCD and active discussion in the current literature 
(see, for instance, 
\cite{paver,Mekhfi:1983az,del,DelFabbro:2002pw,Hussein:2006xr,Kulesza:1999zh,%
Cattaruzza:2005nu,maina,Berger:2009cm,Blok:2010ge,Diehl:2011tt,stir,snig03,%
snig04,snig11,Snigirev:2010ds,Snigirev:2010tk,flesburg,Ryskin:2011kk,%
Bartels:2011qi} and references therein). 

A greater rate of events containing multiple hard interactions is 
anticipated at the LHC  with respect to the experiments mentioned above 
due to the much higher luminosity and greater energy of the LHC. Moreover 
the products from multiple interactions will represent an important 
background~\cite{del,DelFabbro:2002pw,Hussein:2006xr} to signals from the  
Higgs and other interesting processes and certain types of multiple 
interactions will have distinctive  
signature~\cite{Kulesza:1999zh,Cattaruzza:2005nu,maina,Berger:2009cm} 
facilitating a detailed investigation of these processes experimentally.

The main purpose of this letter is to bring attention to another important 
processes: the production of heavy meson pairs through double parton scattering 
which is definitely not taken into consideration in the current theoretical 
estimations~\cite{Baranov1997,Berezhnoy:2011xy}. Here, however, one 
should mention quite a recent paper~\cite{kom}, in which the contribution 
from the double parton scattering to $\J$-pair production has been 
discussed for the first time for the condition of the LHCb experiment.

Let us recall that, with only the assumption of factorization of the two hard 
parton processes $A$ and $B$, the inclusive cross section of a double parton 
scattering process in a hadron collision is written in the following 
form%~\cite{paver,Gaunt:2009re,Cattaruzza:2005nu}
\begin{eqnarray} 
\label{hardAB}
\sigma^{\rm A B}_{\rm DPS} = & &\frac{m}{2} \sum \limits_{i,j,k,l} \int 
\Gamma_{ij}(x_1, x_2; 
{\bf b_1},{\bf b_2}; Q^2_1, Q^2_2)\nonumber\\
& &\times\hat{\sigma}^A_{ik}(x_1, x_1^{'},Q^2_1) 
\hat{\sigma}^B_{jl}(x_2, x_2^{'},Q^2_2)\nonumber\\
&  &\times\Gamma_{kl}(x_1^{'}, x_2^{'}; {\bf b_1} - {\bf b},{\bf b_2} - 
{\bf b}; Q^2_1, Q^2_2)\nonumber\\
& &\times dx_1 dx_2 dx_1^{'} dx_2^{'} d^2b_1 d^2b_2 d^2b,
\end{eqnarray}
where ${\bf b}$ is the usual impact parameter, that is, the distance 
between the centers of colliding (e.g., the beam and the target) hadrons 
in transverse plane. 
$\Gamma_{ij}(x_1, x_2;{\bf b_1},{\bf b_2}; Q^2_1, Q^2_2)$ are the double 
parton distribution functions, depending on the longitudinal momentum 
fractions $x_1$ and $x_2$ and on the  transverse positions ${\bf b_1}$ 
and ${\bf b_2}$ of the two partons undergoing the hard processes $A$ and 
$B$ at the scales $Q_1$ and $Q_2$; $\hat{\sigma}^A_{ik}$ and 
$\hat{\sigma}^B_{jl}$ are the parton-level subprocess cross sections. 
The factor $m/2$ is a consequence of the symmetry with respect to the 
interacting parton species $i$ and $j$: $m=1$ if $A=B$, and $m=2$ otherwise.

It is typically taken that the double parton distribution functions may be 
decomposed in terms of the longitudinal and transverse components as follows:
\begin{eqnarray} 
\label{DxF}
& &\Gamma_{ij}(x_1, x_2;{\bf b_1},{\bf b_2}; Q^2_1, Q^2_2)\nonumber\\
& &= D^{ij}_h(x_1, x_2; Q^2_1, Q^2_2) f({\bf b_1}) f({\bf b_2}),
\end{eqnarray} 
where $f({\bf b_1})$ is supposed to be an universal function for all kind of 
partons with its normalization fixed as
\begin{eqnarray} 
\label{f}
\int f({\bf b_1}) f({\bf b_1 -b})d^2b_1 d^2b = \int T({\bf b})d^2b = 1,
\end{eqnarray} 
and $T({\bf b}) = \int f({\bf b_1}) f({\bf b_1 -b})d^2b_1 $ is the overlap  
function. 

If one makes a further assumption that the longitudinal component
$D^{ij}_h(x_1, x_2; Q^2_1, Q^2_2)$ reduces to a product of two independent 
one parton distributions,
\begin{eqnarray} 
\label{DxD}
D^{ij}_h(x_1, x_2; Q^2_1, Q^2_2) = D^i_h (x_1; Q^2_1) D^j_h (x_2; Q^2_2),
\end{eqnarray}
the cross section of the double parton scattering can be expressed in a 
simple form
\begin{eqnarray} 
\label{doubleAB}
& \sigma^{\rm A B }_{\rm DPS} = \frac{m}{2} \frac{\sigma^{ A}_{\rm SPS} 
\sigma^{ B}_{\rm SPS}} {\sigma_{\rm eff}}, \\
& \sigma_{\rm eff}=[ \int d^2b (T({\bf b}))^2]^{-1}.
\end{eqnarray} 
In this representation and at the factorization of longitudinal and transverse 
components, the inclusive cross section of single hard scattering reads
\begin{eqnarray} 
\label{hardS}
\sigma^{A}_{\rm SPS}
=& &\sum \limits_{i,k} \int D^{i}_h(x_1; Q^2_1) f({\bf b_1})
\hat{\sigma}^A_{ik}(x_1, x_1^{'})\\ 
& &\times D^{k}_{h'}(x_1^{'}; Q^2_1)f( {\bf b_1} - {\bf b}) dx_1 dx_1^{'}  
d^2b_1  d^2b \nonumber
\end{eqnarray}
\begin{eqnarray}
= \sum \limits_{i,k} \int D^{i}_h(x_1; Q^2_1)
\hat{\sigma}^A_{ik}(x_1, x_1^{'}) D^{k}_{h'}(x_1^{'}; Q^2_1) dx_1 
dx_1^{'}.\nonumber
\end{eqnarray}

These simplifying assumptions, though rather customary in the literature 
and quite convenient from a computational point of view, are not 
sufficiently justified and are under the revision 
now~\cite{Blok:2010ge,Diehl:2011tt,stir,flesburg,Ryskin:2011kk}. However, 
the starting cross section formula~(\ref{hardAB}) was found (derived) in 
many works 
(see, e.g., Refs.~\cite{paver,Mekhfi:1983az,Blok:2010ge,Diehl:2011tt,stir})  
in the momentum representation  using the light-cone variables and the 
same approximations as thouse applied to the processes with a single hard 
scattering. 

Nevertheless, we restrict ourselves to this simple form~(\ref{doubleAB}) 
regarding it as the first estimation of the contribution from the double 
parton scattering to the inclusive double heavy meson production. The 
presence of the correlation term in the two-parton distributions results 
in the decrease~\cite{Snigirev:2010tk,flesburg,Ryskin:2011kk} of the 
effective cross section $\sigma_{\rm eff}$ with the growth of the 
resolution scales $Q_1$ and $Q_2$, while the dependence of 
$\sigma_{\rm eff}$ on the total energy at fixed scales is rather 
weak~\cite{flesburg}.  Thus, in fact, we obtain the minimal estimate of 
the contribution under consideration. 

The CDF and D0 measurements give $\sigma_{\rm eff} \simeq$ 15 mb, which is 
roughly 20$\%$ of the total (elastic + inelastic) $p{\bar p}$ cross section 
at the Tevatron energy. We will use this value in our further estimations.

Let us start from the double $\J$ production, since the LHCb Collaboration 
has recently reported a first measurement~\cite{LHCb} of this process
\begin{eqnarray} \label{2Jpsi}
\sigma^{\J\J}=5.6\pm1.1\pm1.2 ~{\rm nb}
\end{eqnarray}
with both $\J$'s in the rapidity region $2<y^{\J}<4.5$ and with the 
transverse momentum $p_T^{\J}<10$ GeV/$c$ in proton-proton collsions at 
a center-of-mass energy of $\sqrt{s}=7$ Tev. Earlier this Collaboration 
has already measured~\cite{lhcb2010} the single inclusive $\J$ production 
cross section within the same kinematical cuts as above 
\begin{eqnarray} \label{Jpsi}
\sigma^{\J}_{\rm SPS}=7.65\pm0.19\pm1.10^{+0.87}_{1.27}~\mu{\rm b}.
\end{eqnarray}

Using Eq. (\ref{doubleAB}) we obtain immediately a simple estimation of 
the contribution from the double parton scattering at the same kinematical 
conditions
\begin{eqnarray} 
\label{2Jpsidps}
\sigma^{\J\J}_{\rm DPS} =\frac{1}{2}\frac{\sigma^{\J}_{\rm SPS}%
\sigma^{\J}_{\rm SPS}}{\sigma_{\rm eff}}\simeq 2.0~{\rm nb}.
\end{eqnarray}
This value is quite comparable with the cross section through the 
``standard'' mechanism of the double $\J$ yield ~\cite{Berezhnoy:2011xy}
\begin{eqnarray} 
\label{2Jpsisps}
\sigma^{\J\J}_{\rm SPS}=4.15 ~{\rm nb},
\end{eqnarray}
and the theoretical prediction for the contribution from both scattering 
modes to the cross section
\begin{eqnarray} 
\label{2Jpsisps+dps}
\sigma^{\J\J}_{\rm SPS}+\sigma^{\J\J}_{\rm DPS}=6.15~{\rm nb},
\end{eqnarray}
is very close to the experimentally observed cross section~(\ref{2Jpsi}) 
of double $\J$ production.

%From our point of view this simple equality matching is the first 
%evidence in favour of the presence of double parton scattering mode in 
%the double $\J$ yield. 
A method to measure the double parton scattering at the LHCb 
using leptonic final states from the decay of two promt $\J$ mesons 
is disccused in Ref.~\cite{kom}.
%%%%%%%%%%%%%
It is worth mentioning on the other hand that the predictions on the 
double $\J$ production are very sensitive to the choice of the 
renormalization scale (because of the ${\cal O}(\alpha_s^4)$ dependence 
of the  $\sigma^{\J\J}_{\rm SPS}$ cross section), and so, the LHCb 
experimental results can also be accommodated by the SPS mechanism alone
(see below).

An even better evidence for the double parton scattering process can be 
found in the production of $\C$ pairs. The production of $P$-wave states 
is suppressed relative to the production of $S$-wave states because of 
the hierarchy of the wave functions 
$|\R_{\J}(0)|^2 \gg |\R'_{\C}(0)|^2/m_{\chi}^2$ 
leading to the inequality
$\sigma^{\J\J}_{\rm SPS} \gg \sigma^{\C\C}_{\rm SPS}$.
Indeed, as one can learn from Fig. 8 in Ref.~\cite{Baranov1997}, the 
inclusive double $\C$ production is suppressed in comparison with the 
inclusive double $\J$ production by more than two orders of magnitude. 

At the same time, the inclusive production of single $\J$ and $\C$
states shows nearly the same rates. The latter property is supported by 
both theoretical \cite{Baranov2002} and experimental \cite{CDF1,CDF2,CDF3} 
results. The reason comes from the fact that the $\C$ mesons are produced 
in a direct $2\to 1$ gluon-gluon fusion $g+g\to\C$, while the $\J$ mesons 
are produced in a $2\to 2$ subprocess $g+g\to\J+g$, where an additional 
final state gluon is required by the color and charge parity conservation. 
As a consequence, the invariant mass of the produced system is typically 
much higher in the $J$ case than in the $\C$ case. (Besides that, the 
structure of the matrix element is such that it vanishes when the 
coproduced gluon becomes soft. This further suppresses the production 
of low-mass states.)

Taken together, the suppression factors coming from the lower wave 
function on the $\C$ side, and from the higher final state mass and 
extra $\alpha_s$ coupling on the $\J$ side nearly compensate each other 
making the inclusive production cross sections comparable in size:
$\sigma^{\C}_{\rm SPS} \simeq \sigma^{\J}_{\rm SPS}$.
As a consequence, we get
$\sigma^{\C\C}_{\rm DPS} \simeq \sigma^{\J\J}_{\rm DPS}$
and $\sigma^{\C\C}_{\rm DPS} \gg \sigma^{\C\C}_{\rm SPS}$.
Thus, if observed, the production of a $\C\C$ pair would yield a clear 
and unambiguous indication of the double parton scattering process.
The need in detecting the decay photon $\C\to\J+\g$ brings sertain 
difficulties in the experimental procedure, but the task seems still 
feasible as the production cross section is not small.

Another tempting possibility is to consider the simultaneous production 
of $\J$ and $\C$. In the SPS mode, this process is forbidden at the 
leading order (LO) by the charge parity conservation but is possible at
the next-to-leading order (NLO), $g+g\to\J+\C+g$. The corresponding 
cross section is then suppressed by one extra power of $\alpha_s$ and 
by the $\C$ wave function. Alternatively, it can proceed via the soft 
final-state gluon radiation (the so called color octet model). The 
estimations of the cross section are then model dependent and rather 
uncertain, but even with the largest acceptable values for the color 
octet matrix elements one arrives at a suppression factor of about two 
orders of magnitude \cite{Baranov1997}. For the DPS mode we still expect 
no suppression, $\sigma^{\J\C}_{\rm DPS}\simeq\sigma^{\J\J}_{\rm DPS}$.

Yet another interesting process is the production of particles from
different flavor families, say, $\J$ and $\Y$ mesons. Once again, this 
process is not possible at the leading order in the SPS mode and only 
can occur either at the NNLO (next-to-next-to-leading order) 
${\cal O}(\alpha_s^6)$, or via the color-octet transitions, or by 
means of the production and decay of $P$-wave mesons (i.e., 
$g+g\to\C+\chi_b$ followed by $\C\to\J+\g$ and $\chi_b\to\Y+\g$). 
So, the SPS mode is always suppressed: either by the extra powers of 
$\alpha_s$, or by the color-octet matrix elements, or by the $P$-state 
wave function, and the DPS mode becomes the absolutely dominant one:
$\sigma_{\rm DPS}^{\J\;\Y}\gg\sigma_{\rm SPS}^{\J\;\Y}$.

Now, to be more precise, we will derive some numerical predictions.
In doing so, we rely upon perturbative QCD and nonrelativistic bound
state formalism \cite{CSM_S,CSM_P} with only the color-singlet channels 
taken into consideration. Also, we accept the $k_t$-factorization ansatz 
\cite{GLR,CCH,CE} for the parton model. The computational technique is 
explained in every detail in Ref. \cite{Baranov2002}, and the parameter 
setting is as follows. The meson masses are taken from the Particle Data 
Book \cite{PDG}, and the heavy quark masses are set equal to one half of 
the respective meson masses; the radial wave functions of $\J$ and $\Y$ 
mesons are supposed to be known from their leptonic decay widths \cite{PDG}
and are set to $|\R_{\J}(0)|^2$ = 0.8 GeV$^3$ 
           and $|\R_{\Y}(0)|^2$ = 6.48 GeV$^3$; 
the wave functions of the $P$-states are taken from the potential model 
\cite{EicQui}, $|\R^\prime_{\chi_c}(0)|^2$ = 0.075 GeV$^5$ 
           and $|\R^\prime_{\chi_b}(0)|^2$ = 1.44 GeV$^5$; 
the renormalization scale in the strong coupling $\alpha_s(\mu^2)$ is 
chosen as the meson transverse mass $\mu^2=m^2+p_t^2$; and we use the 
A0 parameterization from Ref. \cite{A0} for the unintegrated gluon density.

In the present note we will restrict ourselves to the conditions of the
LHCb experiment, since the Collaboration has already recorded the
production of $\J$ pairs. Predictions for other experimental conditions
can be made in an essentially similar way.

Within the theoretical model described above, we get for the direct
inclusive $\J$ production
\begin{equation}
\sigma_{\rm SPS}^{\J}({\rm direct}) = 7.1~\mu{\rm b},
\end{equation}
and for the $\C$ mesons
\begin{equation}
\sigma_{\rm SPS}^{\chi_{c1}} = 1.5~\mu{\rm b}, \quad
\sigma_{\rm SPS}^{\chi_{c2}} = 5.1~\mu{\rm b}.
\end{equation}
After multiplying these numbers by appropriate branching ratios \cite{PDG}
$Br(\chi_{c1}{\to}\J{+}\g)=35\%$ and $Br(\chi_{c2}{\to}\J{+}\g)=20\%$ and 
summing the direct and indirect contributions together we get for the 
prompt $\J$ yield
\begin{eqnarray}
\sigma_{\rm SPS}^{\J} &=& \sigma_{\rm SPS}^{\J}({\rm direct})
                      + \sigma_{\rm SPS}^{\J}({\rm from}~\C)\n \\
   &=& 7.1~\mu{\rm b} + 1.6~\mu{\rm b} =  8.7~\mu{\rm b}.
\end{eqnarray}
This result is in reasonable agreement with the experimental measurement
(9), thus giving support to our theoretical model. Quite similarly, we
get for the $b\bar{b}$ mesons
\begin{equation}
\sigma_{\rm SPS}^{\Y}({\rm direct}) = 140~{\rm nb},
\end{equation}
and
\begin{equation}
\sigma_{\rm SPS}^{\chi_{b1}} = 18~{\rm nb}, \quad
\sigma_{\rm SPS}^{\chi_{b2}} = 91~{\rm nb}.
\end{equation}

Then one can easily obtain for the DPS mode
\begin{eqnarray}
&&\sigma^{\J\J}_{\rm DPS} = 1.7~{\rm nb}, \\
&&\sigma^{\J\J}_{\rm DPS}({\rm both~from}~\C) = 0.9~{\rm nb}, \\
&&\sigma^{\J\Y}_{\rm DPS} = 0.07~{\rm nb}.
\end{eqnarray}
The reader can continue deriving predictions for other DPS combinations.

To calculate the background contribution $\sigma_{\rm SPS}^{\J\J}$ we use
the code developed in \cite{BJ} and extended now \cite{Baranov2011} to 
the \ktf approach:
\begin{equation}
\sigma^{\J\J}_{\rm SPS} = 4~{\rm nb}.
\end{equation}
Variations in the renormalization scale $\mu_R^2$ within a factor of 2
%\marginpar{$\surd$}
around the default value $\mu_R^2 = \hat{s}/4$ make a factor of 1.6 
increasing or decreasing effect on the total production rate.
Employing some different parametrizations for the unintegrated gluon
densities (A+ or A- sets from Ref. \cite{A0}) also changes the predicted
cross section by a factor of 1.6 up or down.
Our central prediction (21) is in reasonable agreement with the data (8). 

The proportion between the visible SPS and DPS contributions can, in 
%\marginpar{$\surd$}
principle, depend on the experimental cuts on the $\J$ transverse 
momentum. However, in the particular case which we are considering 
here, the LHCb Collaboration refers to no cuts on $p_T(\J)$. In fact, 
there are some soft restrictions on the momenta of the decay muons, 
$p_T(\mu)>600$ MeV, but they are taken into account as corrections to 
the acceptance. The final results reported by the Collaboration to
compare with are the acceptance-corrected ones.

It is also worth noting that even in the general case the sensitivity
%\marginpar{$\surd$}
of the ratio $\sigma_{\rm DPS}/\sigma_{\rm SPS}$ to the $p_T$ cuts is
rather weak, because the DPS and SPS contributions show the same $p_T$ 
behavior. This is explained in detail in Ref. \cite{Baranov2011}.
Irrespective of the particular properties of the subprocess matrix 
element, the $p_T$ of the final state is dominated by the transverse 
momentum of the initial gluons, and the individual $\J$ spectra behave
as $1/p_T^4$ in both SPS and DPS modes. 
Moreover, the momenta of the two $\J$ mesons are not correlated. The 
latter is evident in the DPS case and was not a priori evident in the 
SPS case, but turned out to be true (Fig.5 in Ref. \cite{Baranov2011}).
So, the SPS and DPS event topologies are rather similar to each other 
and can hardly be distinguished from one another.

Our calculations agree with the observation made in Ref. \cite{kom} that 
the effects 
of initial parton radiation (that are automatically present in the \ktf 
approach) destroy the original back-to-back $\J\J$ kinematics completely 
washing out the azimuthal correlations. One can potentially distinguish 
the SPS and DPS modes with rapidity correlations, but we anyway find that 
looking at some other meson species is more indicative. In particular, 
the production of $\C\C$, $\J\C$ or $\J\Y$ pairs is totally dominated by 
the DPS mechanism because the SPS mechanism is suppressed for the reasons 
given earlier.

Summing up, we conclude that the processes with pairs of heavy quarkonia
in the final state ($\J\J$, $\C\C$, $\J\C$, $\J\Y$ ) can serve as precise 
probes of the double parton scattering at the LHC and can stimulate
important steps towards understanding the multiparticle QCD dynamics. 

%\section{Summary}

\section*{Acknowledgements}
Discussions with E.E.~Boos, A.I.~Demianov and V.A. Khoze are gratefully acknowledged. 
This work is partly supported by Russian Foundation for Basic Research 
Grants  No. 10-02-93118, the President of Russian Federation for support 
of Leading Scientific Schools Grant No 4142.2010.2. S.B. and N.Z. are 
very grateful to DESY Directorate for the support in the framework of 
Moscow-DESY project  on Monte-Carlo implementation for HERA-LHC, and they 
were supported by RFBR Grant 11-02-01454. A.S. and N.Z. also was supported 
by FASI State contract 02.740.11.0244.

After finishing this work we learned about a similar independent study
%\marginpar{$\surd$}
made by Alexei Novoselov \cite{Novoselov}.

%% The Appendices part is started with the command \appendix;
%% appendix sections are then done as normal sections
%% \appendix

%

\end{document}